\documentclass[twocolumn]{aastex631}

\usepackage{graphicx,amsmath}
\usepackage{amssymb}
\usepackage{nccmath}
\usepackage{upgreek}
\usepackage{subfigure}
\usepackage{color}
\usepackage{bm}
\usepackage{soul}
\usepackage{CJK}
\usepackage{ulem}
\hypersetup{linkcolor=blue, citecolor=blue}

\setcitestyle{notesep={ }}
\newcommand{\Msol}{\hbox{\thinspace $M_{\odot}$}}

\newcommand{\htwo}{H$_2$}
\newcommand{\hi}{H\thinspace{\sc i}}

\newcommand{\ha}{H$\alpha$}

\shortauthors{Dou et al.}

\begin{document}

\title{{\bf \Large From Haloes to Galaxies. IV. The HI reservoir in central spiral galaxies and the implied star formation process}}

\correspondingauthor{Jing Dou, Yingjie Peng, Qiusheng Gu}
\email{doujing@nju.edu.cn, yjpeng@pku.edu.cn, qsgu@nju.edu.cn}

\author[0000-0002-6961-6378]{Jing Dou*}
\affiliation{School of Astronomy and Space Science, Nanjing University, Nanjing 210093, China}
\affiliation{Key Laboratory of Modern Astronomy and Astrophysics (Nanjing University), Ministry of Education, Nanjing 210093, China}

\author{Yingjie Peng*}
\affiliation{Department of Astronomy, School of Physics, Peking University, 5 Yiheyuan Road, Beijing 100871, China}
\affiliation{Kavli Institute for Astronomy and Astrophysics, Peking University, 5 Yiheyuan Road, Beijing 100871, China}

\author[0000-0002-3890-3729]{Qiusheng Gu*}
\affiliation{School of Astronomy and Space Science, Nanjing University, Nanjing 210093, China}
\affiliation{Key Laboratory of Modern Astronomy and Astrophysics (Nanjing University), Ministry of Education, Nanjing 210093, China}

\author[0000-0002-7093-7355]{Alvio Renzini}
\affiliation{INAF - Osservatorio Astronomico di Padova, Vicolo dell'Osservatorio 5, I-35122 Padova, Italy}

\author[0000-0001-6947-5846]{Luis C. Ho}
\affiliation{Kavli Institute for Astronomy and Astrophysics, Peking University, 5 Yiheyuan Road, Beijing 100871, China}
\affiliation{Department of Astronomy, School of Physics, Peking University, 5 Yiheyuan Road, Beijing 100871, China}

\author[0000-0002-4803-2381]{Filippo Mannucci}
\affiliation{Istituto Nazionale di Astrofisica, Osservatorio Astrofisico di Arcetri, Largo Enrico Fermi 5, I-50125 Firenze, Italy}

\author[0000-0002-3331-9590]{Emanuele Daddi}
\affiliation{AIM, CEA, CNRS, Universit\'{e} Paris-Saclay, Universit\'{e} Paris Diderot, Sorbonne Paris Cit\'{e}, F-91191 Gif-sur-Yvette, France}

\author[0000-0001-6469-1582]{Chengpeng Zhang}
\affiliation{George P. and Cynthia Woods Mitchell Institute for Fundamental Physics and Astronomy, Texas A$\&$M University, College Station, TX 77843-4242, USA}
\affiliation{Department of Physics and Astronomy, Texas A\&M University, College Station, TX 77843-4242, USA}

\author[0000-0001-9592-4190]{Jiaxuan Li}
\affiliation{Department of Astrophysical Sciences, 4 Ivy Lane, Princeton University, Princeton, NJ 08544, USA}

\author[0000-0002-8614-6275]{Yong Shi}
\affiliation{School of Astronomy and Space Science, Nanjing University, Nanjing 210093, China}
\affiliation{Key Laboratory of Modern Astronomy and Astrophysics (Nanjing University), Ministry of Education, Nanjing 210093, China}

\author[0000-0002-2504-2421]{Tao Wang}
\affiliation{School of Astronomy and Space Science, Nanjing University, Nanjing 210093, China}
\affiliation{Key Laboratory of Modern Astronomy and Astrophysics (Nanjing University), Ministry of Education, Nanjing 210093, China}

\author{Dingyi Zhao}
\affiliation{Department of Astronomy, School of Physics, Peking University, 5 Yiheyuan Road, Beijing 100871, China}
\affiliation{Kavli Institute for Astronomy and Astrophysics, Peking University, 5 Yiheyuan Road, Beijing 100871, China}

\author[0009-0000-7307-6362]{Cheqiu Lyu}
\affiliation{Department of Astronomy, School of Physics, Peking University, 5 Yiheyuan Road, Beijing 100871, China}
\affiliation{Kavli Institute for Astronomy and Astrophysics, Peking University, 5 Yiheyuan Road, Beijing 100871, China}

\author[0000-0003-3010-7661]{Di Li}
\affiliation{CAS Key Laboratory of FAST, National Astronomical Observatories, Chinese Academy of Sciences, Beijing 100012, China}
\affiliation{School of Astronomy and Space Science, University of Chinese Academy of Sciences, Beijing 100049, China}

\author[0000-0003-3564-6437]{Feng Yuan}
\affiliation{Center for Astronomy and Astrophysics and Department of Physics, Fudan University, Shanghai 200438, People’s Republic of China}

\author[0000-0002-4985-3819]{Roberto Maiolino}
\affiliation{Cavendish Laboratory, University of Cambridge, 19 J. J. Thomson Avenue, Cambridge CB3 0HE, UK}
\affiliation{Kavli Institute for Cosmology, University of Cambridge, Madingley Road, Cambridge CB3 0HA, UK}
\affiliation{Department of Physics and Astronomy, University College London, Gower Street, London WC1E 6BT, UK}

\author[0000-0002-5973-694X]{Yulong Gao}
\affiliation{School of Astronomy and Space Science, Nanjing University, Nanjing 210093, China}
\affiliation{Key Laboratory of Modern Astronomy and Astrophysics (Nanjing University), Ministry of Education, Nanjing 210093, China}

\begin{abstract}

\noindent The cold interstellar medium (ISM) as the raw material for star formation is critical to understanding galaxy evolution. It is generally understood that galaxies stop making stars when, in one way or another, they run out of gas. However, here we provide evidence that central spiral galaxies remain rich in atomic gas even if their star formation rate and molecular gas fraction have dropped significantly compared to ``normal" star-forming galaxies of the same mass. Since \hi\ is sensitive to external processes, here we investigate central spiral galaxies using a combined sample from SDSS, ALFALFA, and xGASS surveys. After proper incompleteness corrections, we find that the key \hi\ scaling relations for central spirals show significant but regular systematic dependence on stellar mass. At any given stellar mass, the \hi\ gas mass fraction is about constant with changing specific star formation rate (sSFR), which suggests that \hi\ reservoir is ubiquitous in central spirals with any star formation status down to $M_* \sim 10^{9}$\Msol. Together with the tight correlation between the molecular gas mass fraction and sSFR for galaxies across a wide range of different properties, it suggests that the decline of SFR of all central spirals in the local universe is due to the halt of \htwo\ supply, though there is plenty of \hi\ gas around. These hence provide critical observations of the dramatically different behavior of the cold multi-phase ISM, and a key to understand the star formation process and quenching mechanism.

\end{abstract}



\section{Introduction} \label{sec:intro} 

The cold interstellar medium (ISM) plays a vital role in driving galaxy formation and evolution. As galaxies evolve, ISM is replenished by gas cooling and accretion from the circumgalactic medium; consumed by star formation; and enriched by stellar winds and feedback \citep[e.g.,][]{2020ARA&A..58..157T,2022ARA&A..60..319S}. The global star formation rates (SFRs) of galaxies are determined by their cold gas content ($M_{\rm gas}$) and star formation efficiency (SFE, defined as SFR/$M_{\rm gas}$) or the depletion timescale (the inverse of SFE, defined as $M_{\rm gas}$/SFR). The SFE is a key galaxy parameter that reflects the efficiency with which a galaxy can convert cold gas into stars. Furthermore, SFE is closely tied to a galaxy’s dynamical time, which represents the time it takes for some fraction of gas to be transformed into stars per galactic orbital time, influenced by the gravitational instability of the cold gas in the galactic disk. This fraction is dependent on the detailed feedback physics \citep{1997ApJ...481..703S,1997ApJ...480..235E,2010MNRAS.407.2091G}.

Observation evidence strongly supports the well-established Kennicutt-Schmidt (KS) type star formation law, asserting a tight correlation between star formation rate (SFR) and total gas content, including atomic and molecular gas \citep[e.g.,][]{1959ApJ...129..243S,1998ApJ...498..541K,2008AJ....136.2846B,2010MNRAS.403..683C,2018Catinella,2012ARA&A..50..531K,2016MNRAS.462.1749S,2022ARA&A..60..319S}. This relationship is more pronounced with molecular gas, especially dense molecular gas \citep[e.g.,][]{2004ApJ...606..271G}, as stars are commonly believed to originate from cold, dense molecular clouds. Numerous studies have corroborated the molecular gas KS law on both global and spatially resolved scales \citep[e.g.,][]{2008AJ....136.2846B,2008Leroy,2020MNRAS.496.4606M,2019ApJ...872...16D,2021ApJ...908...61K}, leading to the proposal of additional scaling relations for molecular gas. These include the SFE of molecular gas (SFE$_{\rm H_2}$ = SFR/$M_{\rm H_2}$) and the molecular gas mass fraction ($\mu_{\rm H_2}$ = $M_{\rm H_2}$/$M_*$), both have been found to tightly depend on the specific star formation rate (sSFR = SFR/$M_*$) \citep[e.g.,][]{2010ApJ...713..686D,2011MNRAS.415...61S,2013ApJ...778....2S,2016MNRAS.462.1749S,2013ApJ...768...74T,2018ApJ...853..179T,2020ARA&A..58..157T,2015ApJ...800...20G,2017A&A...604A..53C,2019A&A...622A.105F,2021ApJ...907..114D,2021ApJ...915...94D,2022ARA&A..60..319S,2023ApJ...943...82S,2023MNRAS.524..923Z}, for both AGNs and non-AGNs \citep[e.g.,][]{2018ApJ...854..158S,2021ApJ...906...38Z,2021ApJS..252...29K,2021A&A...654A.165V,2022A&A...663A..28S}. The molecular gas mass has also been shown to correlate with stellar mass, a relationship often referred to as the molecular gas main sequence \citep[e.g.,][]{2018ApJ...852...74B,2019ApJ...884L..33L}. Although the underlying physical mechanisms of these scaling relations remain debated and somewhat elusive, they are extensively utilized to estimate cold gas masses and to put constraints in theoretical models and simulations.

In \citet{2021ApJ...907..114D}, utilizing the xCOLD GASS survey \citep{Saintonge:2017iz}, we proposed the Fundamental Formation Relation (FFR), a tight relation between sSFR, SFE$_{\rm H_2}$, and $\mu_{\rm H_2}$. Other scaling relations, including the integrated KS law, star-forming main sequence (SFMS), and the molecular gas main sequence can all be derived from this fundamental cube. The molecular gas FFR demonstrates that star formation levels in galaxies are determined by the combined effects of galactic dynamical timescales (related to the gas depletion timescale, 1/SFE) and gas instability (associated with $\mu_{\rm H_2}$). The molecular gas FFR governs the star formation and quenching processes with small scatters. Galaxies with different stellar masses, sizes, structures, metallicities, and in different environments all evolve on the same single scaling relation of $\mu_{\rm H_2}$-SFE$_{\rm H_2}$-sSFR. These unique features and simplicities make the molecular FFR an ideal framework to study galaxy formation and evolution, for instance, to accurately derive the \htwo\ gas cycles in galaxy populations with different stellar masses, from star-forming galaxies to the galaxies in the process of being quenched \citep{2021ApJ...915...94D}.

Another critical component of the ISM in galaxies is atomic gas (\hi). \hi\ generally distributes more extended and loosely bounded comparing to \htwo, and reacts sensitively to external environmental influences like ram pressure stripping and galaxy interactions \citep[e.g.,][]{1984Haynes,2001Solanes,2002ApJ...569..157W,2005Gavazzi,2008AJ....136.2846B,2009AJ....138.1741C,2017A&A...605A..18C,2021ApJ...911...57Z}. Using xGASS survey \citep{2018Catinella}, \citet{2021ApJ...915...94D} explored whether \hi\ follows a similar FFR as \htwo, and found that
the relation between SFE$_{\rm HI}$ (defined as SFR/$M_{\rm HI}$) and sSFR for all galaxies shows significant scatter and strong systematic dependence on the key galaxy properties that have been investigated, revealing that \hi\ does not follow a similar FFR as \htwo. The dramatic difference between \hi\ and \htwo\ indicates that different physical processes, such as environmental effects, notably impact the \hi\ gas, while \htwo\ relations remain insusceptible.

Given that \hi\ serves as the precursor for \htwo, it play a critical role in star formation. To accurately assess the intrinsic properties of \hi\ and its actual role in star formation, we must exclude any external environmental factors to which \hi\ is particularly sensitive. Central galaxies are the most massive galaxies within a dark matter halo and usually locate at the center of a galaxy cluster or group. They are typically distinguished from satellite galaxies, which orbit around the central galaxy. Since external environmental effect mainly operates on satellite galaxies \citep[e.g.,][]{Peng:2010gn,2012ApJ...757....4P}, in this work, only central galaxies are selected to minimize the environmental effects, which allows us to focus on star formation and quenching processes due to internal mechanisms. 

Different galaxy types can follow very different evolutionary paths. We only focus on spiral galaxies for the following reasons. Firstly, many ellipticals nowadays are already quenched at higher redshift \citep[e.g.,][]{2005ApJ...626..680D,2012ApJ...755...26O}. Including them can introduce severe progenitor bias. Secondly, ellipticals in the local universe are mostly transformed from the spiral galaxies via mergers rather than internal secular evolution \citep[e.g.,][]{1992ApJ...393..484B,2005A&A...437...69B,2019MNRAS.486.1404D}. Internal quenching mechanisms, such as AGN feedback alone, cannot directly transform spirals to ellipticals. And also, the formation mechanisms for irregulars and S0 galaxies are still not fully understood and may involve recent mergers or strong interactions as well \citep[e.g.,][]{2011MNRAS.415.1783B,2017A&A...604A.105T,2018ApJ...862..100G}.

Central spiral galaxies are ideal candidates for studying the overall properties and behavior of \hi\ gas and star formation, significantly contributing to our understanding of galaxy formation and evolution. In combination with the molecular FFR, we aim to provide a comprehensive understanding of the cold multi-phase ISM, star formation process and (internally-driven) quenching mechanisms in local galaxies. Throughout this work, we adopt the cosmological parameters: $\Omega_m=0.3, \ \Omega_\Lambda=0.7, \ H_0=70\,\rm {km\,s^{-1} Mpc^{-1}}$.


\section{Sample} \label{sec:sample}


\subsection{The ALFALFA-SDSS matched sample} \label{matched sample}

The main \hi\ sample used in this work is Arecibo Legacy Fast ALFA (ALFALFA) survey \citep{2011AJ....142..170H,2018ApJ...861...49H}, which is a blind \hi-selected survey out to $z \sim$ 0.06. The survey utilized the seven-horn Arecibo L-band Feed Array (ALFA) to map $\sim$ 7000 deg$^2$ of the high Galactic latitude sky in drift-scan mode \citep{2005Giovanelli}. Conducted between 2005 and 2011, ALFALFA covered a frequency range from 1335 to 1435 MHz, corresponding to heliocentric velocities from -2000 km/s to 18000 km/s, and detected $\sim$31,500 extragalactic \hi\ 21cm line sources.

The ALFALFA survey contains sources of quality code 1 and quality code 2. Code 1 refers to the \hi\ sources with the highest quality, evaluated using several criteria: (1) a good signal consistence between the two independent polarizations observed by ALFALFA, (2) a spatial extent consistent with or larger than the telescope beam, (3) the spectral profile is free from radio-frequency interference (RFI), (4) an approximate minimum signal-to-noise ratio (S/N) threshold of 6.5. Code 2 refers to the sources with low S/N ($\lesssim$ 6.5) but have been matched with optical counterparts with known optical redshifts coincident with those measured in \hi. They are also included in the ALFALFA catalog as they are highly likely to be real. Both Code 1 and Code 2 sources are considered \hi\ detections in our analysis.

The parent SDSS Data Release 7 (DR7) \citep{Abazajian:2009ef} sample was retrieved from the SDSS CasJobs site, covering $\sim$9,500 deg$^2$ of the sky. The SDSS spectroscopic survey covers a wavelength range of 3,800 to 9,200$\AA$, utilizing fibers with a diameter of 3 arcseconds on the sky. Following the criteria by \citet{2006MNRAS.373..469B}, galaxies with clean photometry and Petrosian r magnitudes in the range of 10.0 - 18.0 after correction for Milky-Way galactic extinction are selected. The bright limit of 10 magnitudes is set to exclude extremely bright, large nearby galaxies, while the faint limit of 18 magnitudes ensures the completeness of the SDSS Spectroscopic Survey. After removing duplicates, the parent photometric sample contains 1,579,314 objects, of which 72,697 have reliable spectroscopic redshift measurements in the redshift range 0.02 $< z <$ 0.05. This narrow and low redshift range is used to match the depth of the ALFALFA survey. The physical scales within this redshift range vary from 1.2kpc to 2.9kpc.

There are 43,448 SDSS galaxies with reliable spectroscopic redshift measurements in the SDSS and ALFALFA overlapped region within the redshift range of $z =$ 0.02 - 0.05. The ALFALFA \hi\ detections are then cross-matched with the parent SDSS sample in this narrow redshift range. Briefly, the spatial separation between the most probable optical counterpart of each \hi\ detection and SDSS galaxy is less than 5$''$. Also, the velocity difference between the \hi\ source and SDSS galaxy is less than 300 km/s. For most galaxies, the line-of-sight velocity difference is generally less than 100 km/s, and using 300 km/s can encompass more than 99\% of galaxies. Slightly adjusting the matching conditions will not significantly affect the matched sample. It should be noted that the angular resolution is $\sim$ 3.5 arcmin in ALFALFA. Close companions may contaminate the measured \hi\ spectra within the large Arecibo beam. We hence exclude galaxies that have multiple SDSS counterparts within the ALFALFA beam size ($\sim$ 3.5 arcmin) and within a velocity difference of three times the \hi\ line width ($W_{50}$), where $W_{50}$ is defined as the difference between the velocities corresponding to the fitted polynomial at a level of 50\% of the maximum value of the flux on each horn. The threshold of three times the $W_{50}$ is an empirically standard that has been found effective and employed in many studies because it balances the need for accurate matches without missing potentially relevant sources \citep[e.g.,][]{2018ApJ...861...49H}. Also, galaxies that are in dense regions or have close companions are more vulnerable to environmental effect. Since our goal of this work is to investigate in-situ star formation process and mass-quenching (i.e. the internal quenching) mechanism, we try to exclude any environmental effect. About 12\% of ALFALFA galaxies were excluded to avoid contaminations from the interlopers. This subset of \hi\ detections with clean optical counterparts is referred to as the ALFALFA-SDSS matched sample, consisting 9571 galaxies. 

The aperture is also an important issue. The \hi\ spatial scale is, on average, much larger than the optical and \htwo\ scales. Before the Square Kilometre Array \citep[SKA;][]{2009IEEEP..97.1482D}, it is difficult to have a large sample with deep and resolved \hi\ measurements. Nevertheless, with the well-established \hi\ mass-size relation based on resolved deeper \hi\ observation of nearby spiral galaxies \citep[e.g.,][]{1997A&A...324..877B,2016MNRAS.460.2143W}, we can roughly estimate the \hi\ spatial scales of our galaxies. The massive star-forming galaxies (which are also about the most \hi-rich galaxies, in terms of the absolute \hi\ gas mass) on average have a \hi\ gas mass of $\sim 10^{10}\Msol$, which corresponds to a \hi\ diameter of about 50 kpc. The $\sim$ 3.5 arcmin’s beam of the Arecibo telescope corresponds to an angular size of approximately 85 kpc at a redshift of $z =$ 0.02 and 200 kpc at a redshift of $z =$ 0.05, which are larger than the \hi\ diameter. Hence, this should not cause a significant bias. The most relevant bias might be the interlopers into the large beam, as discussed above.


\subsection{Incompleteness corrections} \label{vmax correction}

As detailed in Appendix \ref{appendix a} and illustrated in Figure \ref{dete_limit}, both SDSS and ALFALFA are flux-limited samples, resulting in significant selection biases in both stellar mass and \hi\ gas mass, even within the narrow redshift range of $z =$ 0.02 - 0.05. Therefore, when conducting statistical studies using the ALFALFA-SDSS matched sample, it is essential to apply a joint incompleteness correction to account for these biases.

Following the method introduced in \citet{2019ApJ...884L..52Z}, we have used a joint $V_{\rm max}$ correction to account for the volume incompleteness within the given redshift range. For the ALFALFA-SDSS matched sample, we calculated for each individual galaxy the maximum redshift to which the galaxy can still be observed according to the ALFALFA and SDSS detection limits. This maximum redshift was then used to calculate the maximum observable comoving volume ($V_{\rm max}$) for each galaxy. By assuming that the spatial distribution of our sample is homogeneous and there is no evolution in the comoving volume (within a narrow redshift range), we weight each galaxy using the value of $V_{\rm total}$/$V_{\rm max}$ to account for the galaxies missed in the surveys, where $V_{\rm total}$ is the total comoving volume that our sample spans. The corrections of the SDSS and ALFALFA sample are performed independently and they are combined together to correct the ALFALFA-SDSS matched sample, i.e., we have simultaneously corrected the strong sample selection effects shown in both panels in Figure \ref{dete_limit}.

One of the advantages of this method is that the weighting factor has been calculated for each individual galaxy in our sample. This allows us to perform statistical analysis conveniently for a subsample, e.g., central spirals, ellipticals and uncertains. As we discussed and emphasized in Figure \ref{det_bt} and Appendix \ref{spiral definitions}, galaxies with different visual morphology (i.e., spiral, uncertain and elliptical) have intrinsically distinct \hi\ properties, and should be treated and studied separately, for instance, in a stacking analysis.


\subsection{XGASS}

To test our incompleteness correction on the large but shallow ALFALFA-SDSS sample and verify the results, we have included in our analysis the \hi\ observations from a much deeper \hi\ target survey, the extended GALEX Arecibo SDSS Survey \citep[xGASS;][]{2018Catinella}. It provides the \hi\ gas measurements for 1179 galaxies in the nearby universe. These galaxies are selected only by redshift ($0.01 < z < 0.05$) and stellar mass ($10^{9}\Msol< M_* <10^{11.5}\Msol$) from the overlapped area of the SDSS DR7 spectroscopic survey, the GALEX Medium Imaging Survey \citep{2005ApJ...619L...1M} and projected ALFALFA footprints \citep{2011AJ....142..170H}. Those galaxies with reliable \hi\ detections already available from the 40\% ALFALFA catalog or the Cornell \hi\ digital archive \citep{2005ApJS..160..149S} were not reobserved in xGASS to optimize the efficiency. The rest galaxies were observed with the Arecibo telescope until the \hi\ line was detected, or a limit of a few percent in $M_{\rm HI}/M_*$ was reached. This limit of $M_{\rm HI}/M_*$ is 2\% for galaxies with log $M_*/\Msol >$ 9.7, and a constant gas mass limit of log $M_{\rm HI}/\Msol $= 8 for galaxies with lower stellar masses. Each galaxy is weighted by a correction factor to account for selection effects in stellar mass, described in \citet{2018Catinella}.


\subsection{The selection of central spiral galaxies} \label{central}

It is well known that \hi\ is sensitive to external environmental effects \citep[e.g.,][]{2005Gavazzi,2013MNRAS.436...34C}. Therefore, if we want to focus on the star formation and quenching processes due to internal mechanisms, only central galaxies are selected to minimize the environmental effects that are strongly operating on satellites \citep[e.g.,][]{2021PASA...38...35C,2023MNRAS.523.1268W}.

The galaxies are classified into central and satellite galaxies using the SDSS DR7 group catalogue from \citet{Yang:2007}. The central galaxies are defined as the most massive and most luminous ones in the $r$-band within a given group. ``Centrals with satellites" and ``isolated centrals/singletons" are both included as ``centrals" in our analysis. Among the 4470 central spiral \hi\ detections in ALFALFA, 87\% of them are singletons in \citet{Yang:2007} group catalog, i.e., groups with the richness N = 1. The remaining 13\% are centrals with N $>$ 1 (i.e., with at least one satellite). This is expected since the majority of low-mass centrals (much more abundant than the massive ones) are singletons just above the detection limit, and their fainter satellites are not observed for a flux-limited survey. We have also checked the \hi\ detection fraction and \hi\ scaling relations for ``Centrals with satellites" and ``isolated centrals/singletons". The results show little difference for these two cases and our conclusions remain robust.

As mentioned in the introduction, our aim is to better understand the processes of in-situ star formation and internal-origin quenching. Environmental effects can strongly impact \hi\ gas, hence our analysis focuses exclusively on central spiral galaxies. In practice, spiral galaxies can be classified based on various criteria, such as visual morphology, structural parameters, or kinematics. Each classification method results in significantly different samples, as illustrated in Appendix \ref{spiral definitions} and Figure \ref{structure}. Visually-classified ``spiral/disk" is the most effective parameter to distinguish \hi-rich galaxies and \hi-poor galaxies. This is because \hi\ gas disk (extending to much outer regions than the stellar disk) is most sensitive to external interaction or perturbation, and then followed by the visual morphology of the disk, and the least sensitive parameters are the structural parameters (such as bulge-to-total ratio (B/T), concentration index ($R_{90}$/$R_{50}$) and Sérsic index. Hence, we focus only on central galaxies that are visually defined as spiral galaxies. 

Galaxies are classified into different morphologies using data from the Galaxy Zoo (GZ) project \citep{2011MNRAS.410..166L}. In this project, hundreds of thousands of volunteers are assisted to view each SDSS image and classify the galaxy morphology. A clean sample has been defined by requiring at least 80 \% of the corrected vote to be in a particular category. A morphology flag (``spiral", ``elliptical" or ``uncertain") is assigned to each galaxy after performing a careful debiasing process. The contamination of lenticular or S0s into the GZ clean spiral sample is small, about 3\%\ \citep{2008MNRAS.389.1179L,2009MNRAS.393.1324B,2011MNRAS.410..166L}. Most S0 galaxies with smooth and rounded profiles are classified in GZ as ``elliptical” or ``uncertain”. It should be noted that in GZ, ``spiral" includes disky galaxies with spiral arms and also those without clear spiral arms. In this work, we categorize them both as ``spirals". 

Since galaxy mergers can have a complicated effect on the star formation of galaxies by enhancing or suppressing star formation, depending on the gas content of the merging galaxies and also the phase of the merger, we exclude the merger systems in our sample if the vote fractions of a merger are greater than 30\%. As we will show later in Figure \ref{vmax} and \ref{det_bt}, central spiral galaxies defined by visual morphology classified by the GZ project achieve a very high \hi\ detection fraction, which is essential for obtaining unbiased intrinsic \hi\ scaling relations. In contrast, central spiral galaxies defined by structural parameters show a significantly lower average \hi\ detection fraction.


\subsection{Stellar mass and SFR measurements} \label{M and sfr}

The stellar masses ($M_*$) of the galaxies are estimated from the $k-correction$ program \citep{2007AJ....133..734B} with \citet{2003MNRAS.344.1000B} stellar population synthesis model and a \citet{2003PASP..115..763C} initial mass function (IMF), which show a small scatter of $\sim$ 0.1 dex compared with the published stellar masses of \citet{2003MNRAS.341...33K}. 

The star formation rates (SFRs) are taken from the value-added MPA-JHU catalog \citep{2004MNRAS.351.1151B}, which are based on the \ha\ emission line luminosities. These luminosities are corrected for extinction using the H$\alpha$/H$\beta$ ratio. To correct for the aperture effects, the SFRs outside the SDSS 3$''$ fiber were obtained by performing the spectral energy distribution fitting to the $ugriz$ photometry outside the fiber, using the models and methods described in \citet{2007ApJS..173..267S}. The \ha\ emission for AGN and composite galaxies can be contaminated by central nuclear activities. Their SFRs are derived based on the strength of 4000\textup{~\AA} break, calibrated with H$\alpha$ for non-AGN, pure star-forming galaxies \citep[see details in][]{2004MNRAS.351.1151B}. These SFRs are computed for a Kroupa IMF and we convert them to a Chabrier IMF using log SFR (Chabrier) = log SFR (Kroupa) - 0.04. Different SFR estimators may produce different results. The results using different SFR estimators are shown and discussed in the Appendix \ref{sfr indicators}.


\section{Results}


\subsection{HI detection fraction}

\begin{figure*}[htbp]
    \begin{center}
       \includegraphics[width=180mm]{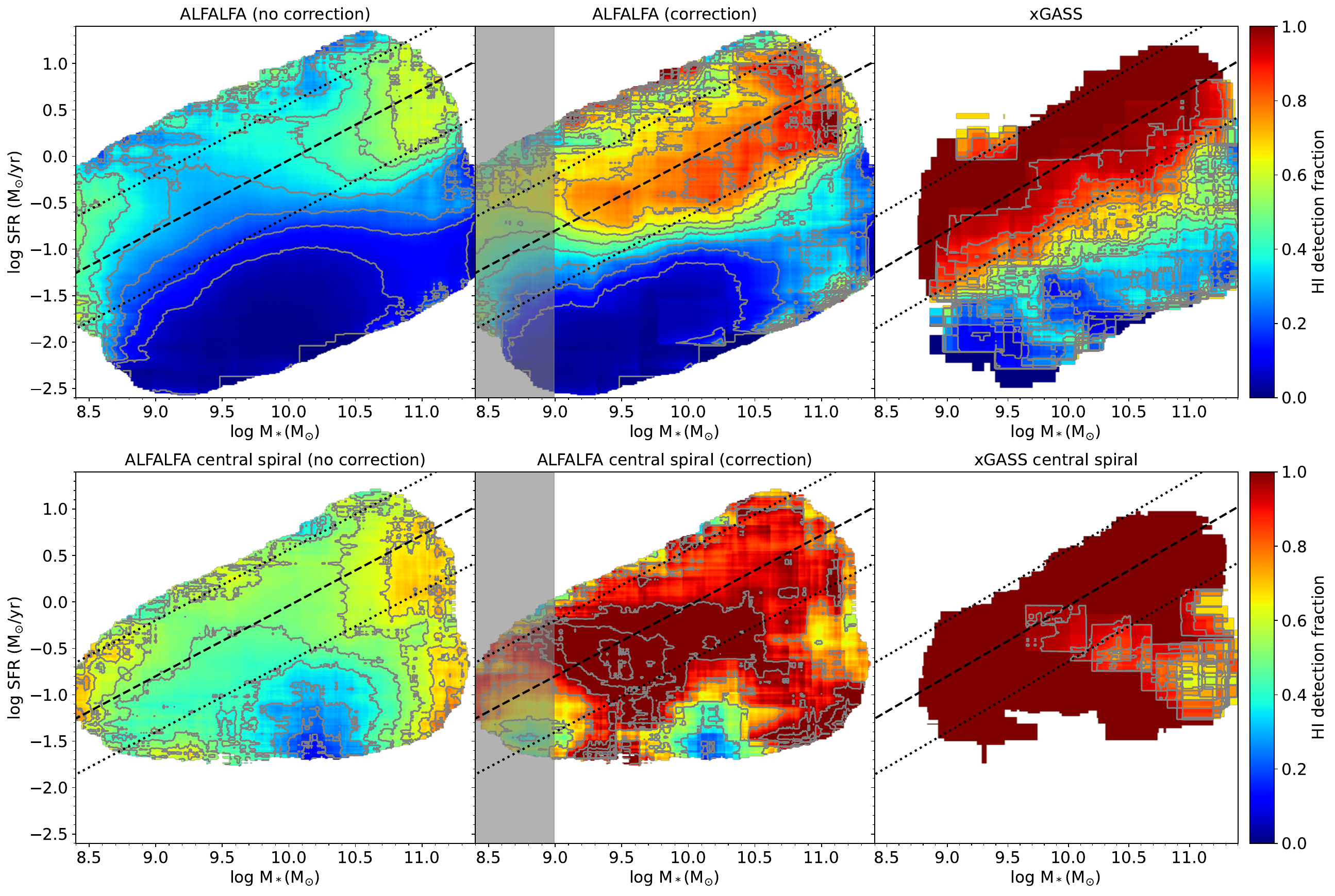}
    \end{center}
\caption{\hi\ detection fraction as a function of stellar mass and star formation rate for different samples. These are obtained by using a moving box of size 0.5 dex in SFR and 0.5 dex in $M_*$. In all panels, the dashed line indicates the position of the star-forming main sequence defined in \citet{2015Renzini}. The dotted lines indicate $\pm$0.4 dex scatter around the main sequence. Contour lines connect points that have the same detection fraction. The grey shades in the middle panels indicate the region with $M_* < 10^{9}\Msol$, where the sample becomes incomplete even with corrections. Left: \hi\ detection fraction in the ALFALFA-SDSS matched sample for all galaxies (upper left panel) and central spiral galaxies (lower left panel) without incompleteness corrections. Middle: The same as left panels, but with incompleteness corrections. Right: \hi\ detection fraction in the xGASS sample for all galaxies (upper right panel) and central spiral galaxies defined by visual morphology (lower right panel).}
 \label{vmax}
\end{figure*}

\begin{figure*}[htbp]
    \begin{center}
       \includegraphics[width=180mm]{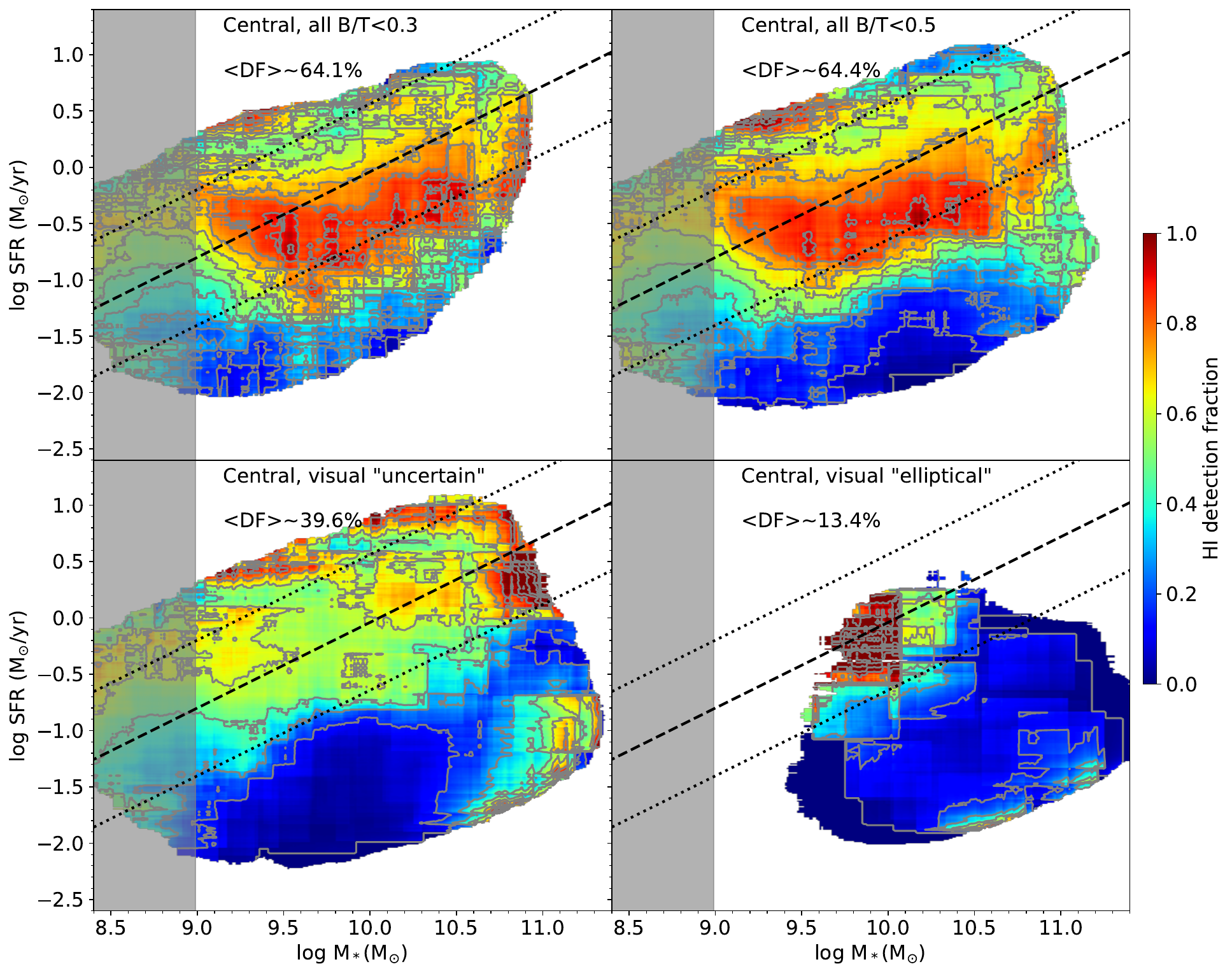}
    \end{center}
\caption{As for Figure \ref{vmax}, using the ALFALFA-SDSS matched sample with incompleteness correction. The upper two panels are for all central galaxies with B/T $<$ 0.3 (upper left panel) and B/T $<$ 0.5 (upper right panel). The lower two panels are for central, visually-defined ``uncertain" galaxies (lower left panel) and central, visually-defined ``elliptical" galaxies (lower right panel) from the Galaxy Zoo project. In each panel, the average \hi\ detection fraction for the given subsample above the completeness mass of $M_* \sim 10^{9}\Msol$ is given in the legend.}
 \label{det_bt}
\end{figure*}

ALFALFA is a blind, relatively shallow, flux-limited \hi\ survey over a significant volume. As discussed in Section \ref{vmax correction} and shown in Figure \ref{dete_limit}, even within the narrow redshift range of $z =$ 0.02 - 0.05, both ALFALFA and SDSS are highly biased samples, with strong selections in \hi\ gas mass, \hi\ line width and redshift for ALFALFA, and in stellar mass, color and redshift for SDSS. Hence, when combining ALFALFA and SDSS (i.e., the ALFALFA-SDSS matched sample), a joint incompleteness correction is necessary, as discussed in Section \ref{vmax correction}. The corrected ALFALFA-SDSS matched sample contains 9571 reliable \hi\ detections, and 4470 of them are central spirals classified by visual morphology.

In Figure \ref{vmax}, we show the \hi\ detection fraction (DF) on the SFR-$M_{*}$ plane for different samples with or without the incompleteness correction. The \hi\ DF is defined as the ratio between the number of galaxies with \hi\ detections and the number of all galaxies in the parent SDSS sample, obtained by using a moving box of size 0.5 dex in SFR and 0.5 dex in $M_*$.

The upper left panel of Figure \ref{vmax} shows the raw \hi\ DF for all galaxies in the ALFALFA-SDSS matched sample without the incompleteness correction. The overall \hi\ DF is evidently low, even for galaxies on the SFMS. The lower left panel shows the raw \hi\ DF for central spirals (defined by the visual morphology as in the Galaxy Zoo project; see Section \ref{central} for details). Compared to all galaxies, the DF increases significantly, particularly for those with low SFRs.

The middle panels are similar to the left panels but have done incompleteness corrections. In both panels, the corrected DFs are significantly elevated compared to the raw DFs shown in the left two panels. For the case of all galaxies (upper middle panel), the corrected average \hi\ DF for galaxies on and above the SFMS (above the lower dotted lines) is significantly higher than that below the SFMS (below the lower dotted lines), indicating that quenching or quenched galaxies are indeed cold gas-poor, on average. Remarkably, in the case of central spiral galaxies (lower middle panel), the corrected \hi\ DF is further enhanced. Although it should be noted that below the SFMS, the DF is slightly lower in both ALFALFA and xGASS samples. The average \hi\ DF for central spiral galaxies, including low-mass galaxies with $M_{*} > 10^{9}$\Msol\ and galaxies below the SFMS, is $\sim$ 95\%, i.e., there is a ubiquitous \hi\ reservoir in the vast majority of central spirals. 

To test our incompleteness correction to the ALFALFA-SDSS sample and the robustness of our results, we show the results obtained from a much deeper \hi\ target survey xGASS in the right two panels. There are 662 reliable \hi~detections (i.e., the \hi~line is detected and not confused by close companions) in the xGASS sample, and 273 of them are central spirals (classified by visual morphology). The \hi\ DFs for all galaxies and central spiral galaxies in the xGASS sample (right two panels) look distinct from the left two panels (which have not done incompleteness corrections) but similar to the middle two panels (which have done proper incompleteness corrections). In particular, the xGASS central spirals consistently show a nearly 100\% \hi\ DF across a wide range of stellar mass and SFR. The overall raw \hi\ DF is $\sim$ 96\%, and $\sim$ 98\% with the statistical weights given by \citet{2018Catinella}, and hence strongly supports the ubiquitous \hi\ gas content in central spiral galaxies. There are some ``green holes” representing a lower DF in both the ALFALFA central spirals and xGASS central spirals panels. They do not seem to follow a systematic trend, and they do not occupy a similar region on the SFR-$M_*$ plane. These could be caused by small number statistics (for the small but deep xGASS) or imperfect incompleteness corrections (for the large but shallow ALFALFA). Future larger and deeper surveys will verify these results.

As shown in Figure \ref{structure}, at a given stellar mass, a significant amount of visually-defined central ``spirals" have large values of $R_{90}$/$R_{50}$, B/T and Sérsic index, and will be classified as ``elliptical" if the definition is in accordance with their structural parameters. As shown in \citet{2021ApJ...911...57Z}, these central spirals with a massive bulge have lower sSFR, but are \hi-rich. On the other hand, as shown in Figure \ref{structure},  many visually-defined ``uncertain" galaxies have small values of $R_{90}$/$R_{50}$, B/T and Sérsic index, and will be classified as ``spiral" according to their structural parameters. These ``uncertain" galaxies, on average, are \hi-poor as shown in the lower left panel in Figure \ref{det_bt}, in particular for those below the SFMS, which are almost all \hi\ non-detections. Therefore, using ``spiral" defined by quantitative structural parameters produces a lower \hi\ DF towards low sSFR (by including many \hi-poor ``uncertain"). In the upper panels of Figure \ref{det_bt}, we show that for central galaxies with B/T $<$ 0.3 and B/T $<$ 0.5, the \hi\ DFs drop significantly towards low sSFR in both cases, consistent with the results of \citet{2020MNRAS.494L..42C}; and also the overall \hi\ DF is $\sim$ 64\% for both B/T cuts, much lower than that of the central spirals defined by visual morphology ($\sim$ 95\%).

To complete, we also show the \hi\ DFs for the central ``elliptical" galaxies defined from the Galaxy Zoo project in the lower right panel in Figure \ref{det_bt}. The overall \hi\ DF is only $\sim$ 13.4\%, consistent with the fact that ellipticals, on average, are cold gas-poor systems. These results show that galaxies with different visual morphology (i.e., spiral, uncertain and elliptical) have intrinsically distinct \hi\ properties, and should be treated and studies separately, for instance, in stacking analysis.


\subsection{HI scaling relations}

\begin{figure*}[htbp]
    \begin{center}
       \includegraphics[width=180mm]{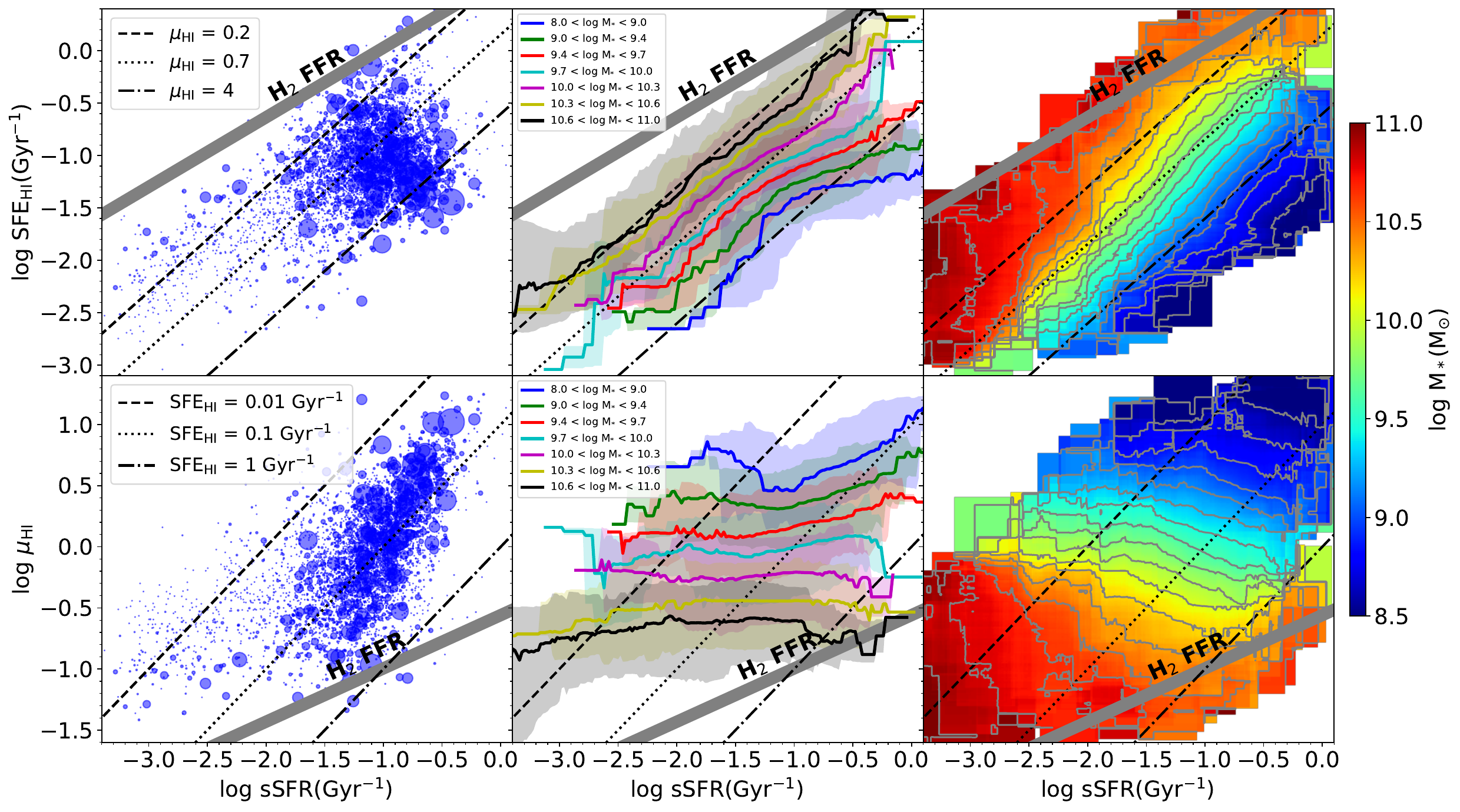}
    \end{center}
\caption{The SFE$_{\rm HI}$-sSFR relation (upper panels) and $\mu_{\rm HI}$-sSFR relation (lower panels) of central spiral galaxies in ALFALFA-SDSS matched sample with incompleteness corrections. Left: The distribution of galaxies on the SFE$_{\rm HI}$-sSFR and $\mu_{\rm HI}$-sSFR planes. The size of each point represents the weight used for incompleteness correction. Middle: The median SFE$_{\rm HI}$ and $\mu_{\rm HI}$ as a function of sSFR in different stellar mass bins, determined by using moving median with a sliding window of 0.5 dex in sSFR. Error bars on each line indicate the 5\% and 95\% percentile of the distribution. As discussed in Appendix \ref{bias} and Figure \ref{map test}, these 1-dimensional analysis can be heavily affected by the shape of the sample distribution and lead to biased results. Right: Average stellar mass on the SFE$_{\rm HI}$-sSFR and $\mu_{\rm HI}$-sSFR planes, calculated with a moving box of 0.5 dex in both sSFR and SFE$_{\rm HI}$ (or $\mu_{\rm HI}$). The contour lines mark the equal stellar mass distribution. In each panel, the black diagonal lines indicate three different values of constant $\mu_{\rm HI}$ of 0.2, 0.7 and 4 (upper panels), or SFE$_{\rm HI}$ of 0.01, 0.1 and 1 Gyr$^{-1}$ (lower panels). The grey thick lines indicate the fundamental formation relation (FFR) of molecular gas found in \citet{2021ApJ...907..114D}, corresponding to $\mu_{\rm H_2}$ on the y-axis.}
 \label{3d mpa}
\end{figure*}

The high overall \hi\ DF of $\sim$ 95\% for the central spiral galaxies makes it possible to derive the unbiased \hi\ scaling relations for this particular population. The left panels of Figure \ref{3d mpa} show the distributions of the central spirals in the ALFALFA-SDSS matched sample on the SFE$_{\rm HI}$-sSFR plane (upper left panel, where \hi\ star formation efficiency SFE$_{\rm HI}$ = SFR/M$_{\rm HI}$) and $\mu_{\rm HI}$-sSFR plane (lower left panel, where \hi\ gas fraction $\mu_{\rm HI} = M_{\rm HI} / M_*$). The SFE$_{\rm HI}$ is the inverse of the \hi\ gas depletion timescale, describing how efficiently the galaxy can convert the available cold \hi\ gas into stars, or how long the \hi\ gas reservoir would be depleted by current star formation activities. Although SFR and M$_{\rm HI}$ happen at very different spatial scales, SFE$_{\rm HI}$ can be interpreted as the product of the star formation efficiency of \htwo\ and the molecular-to-atomic gas mass ratio $M_{\rm H_2}/M_{\rm HI}$ (SFE$_{\rm HI}$ = SFE$_{\rm H_2} \times M_{\rm H_2}/M_{\rm HI}$). The size of each dot represents the weight used for incompleteness correction. The weighting factor for most galaxies ranges from 1 to 10, with the median value is about 1.34 and the mean value is about 2.73, while the maximum value is around 42. Different studies may use different sSFR cuts to define star-forming, green valley, and passive galaxies. Typically, star-forming galaxies are defined as log sSFR (Gyr$^{-1}$) $>$ - 1.5, fully quenched galaxies are defined as log sSFR (Gyr$^{-1}$) $<$ -2.5, with green valley galaxies occupying the range in between. It is evident that even for star-forming galaxies with high sSFR, many of them have a large dot size, i.e., a large weighting factor, suggesting that the sample there is highly incomplete. Therefore, the incompleteness correction is necessary for both star-forming and passive galaxies. On average, both SFE$_{\rm HI}$ and $\mu_{\rm HI}$ increase with increasing sSFR, but with significant scatters. 

The thick grey line in each panel in Figure \ref{3d mpa} shows the fundamental formation relation (FFR) for \htwo\ as found in \citet{2021ApJ...907..114D}. As discussed in \citet{2021ApJ...907..114D}, for \htwo, both SFE$_{\rm H_2}$-sSFR relation and $\mu_{\rm H_2}$-sSFR relation are very tight, independent of other key galaxy parameters, and their scatters can be entirely explained by measurement errors. However, the \hi\ relations for all galaxies does not follow a similar tight FFR as \htwo\ \citep{2021ApJ...915...94D}. Here we further show that even for the central spiral galaxies, their \hi\ relations still exhibit significant scatters as shown in Figure \ref{3d mpa}. It is well known that SFR does not correlate well with \hi\ content while it does so with \htwo \citep{2009ApJ...699..850K,2016MNRAS.462.1749S,2012ARA&A..50..531K,2021ApJ...908...61K}. In a sense, here we demonstrate that this lack of correlation is due to central spirals, which have \hi\ even when their SFR is low.

The middle panels in Figure \ref{3d mpa} use the common 1-dimensional analysis to show the median SFE$_{\rm HI}$ and $\mu_{\rm HI}$ as a function of sSFR for different stellar mass bins, calculated with a sliding window of 0.5 dex in sSFR, which is chosen to match the typical observational uncertainty in measuring sSFR and also to include a sufficient number of data points, ensuring statistical robustness. Error bars on each line indicate the 5\% and 95\% percentiles of the distribution. The right panels use the 2-dimensional analysis to show the average $M_*$ on the SFE$_{\rm HI}$-sSFR plane and $\mu_{\rm HI}$-sSFR plane, calculated with a moving box of 0.5 dex in sSFR, SFE$_{\rm HI}$ and $\mu_{\rm HI}$, and the contour lines mark the equal stellar mass distribution. The overall trends shown in the corresponding panels are similar, but with notable differences in the low stellar mass regions. As discussed in detail in Figure \ref{map test}, the widely used 1-dimensional analysis (which considers only the variation of y at a fixed x) can be heavily affected by the shape of the sample distribution, in particular for the low stellar mass bin; while the 2-dimensional analysis (which takes the variation of both the x- and y-axes into account) is more robust. Therefore, the results shown in the right panels in Figure \ref{3d mpa} are more objective and less biased representations of the true underlying trends.

The contour lines in the upper right panel in Figure \ref{3d mpa} are almost all parallel to the constant $\mu_{\rm HI}$, which is consistent with those shown in Figure \ref{map test}, where the distribution of the galaxies (indicated by the ridge of the contour lines) is around the constant $\mu_{\rm HI}$ and the slope of the distribution (indicated by the red line) is about unity at a given stellar mass. Consistent results are also shown in the lower right panel of Figure \ref{3d mpa}, where the contour lines are about horizontal. It is expected, since by definition, sSFR = $\mu_{\rm HI} \times$ SFE$_{\rm HI}$; hence the $\mu_{\rm HI}$-sSFR relation is closely related to the SFE$_{\rm HI}$-sSFR relation. Both panels indicate that for central spirals, at a given stellar mass, star-forming galaxies, green valley galaxies and passive galaxies all host a similar amount of \hi\ gas. The \hi\ gas fraction primarily depends on stellar mass and weakly depends on sSFR. We also investigate the SFE$_{\rm HI}$-sSFR and the $\mu_{\rm HI}$-sSFR relation for central spirals in the xGASS sample, which exhibit consistent general trends but with considerable scatter compared to ALFALFA due to their much smaller sample size. 

We also tested the SFE$_{\rm HI}$-sSFR and the $\mu_{\rm HI}$-sSFR relation for central spirals defined by other spiral definitions, such as concentration, B/T, sersic index and probability of a galaxy to be late-type disk galaxy with T-type $>$ 0 as shown in Figure \ref{structure} in the Appendix \ref{spiral definitions}. Interestingly, these different definitions do not significantly affect these scaling relations. This is because ALFALFA is a relatively shallow survey and these relations are derived using only galaxies with \hi\ detections. However, as demonstrated in Figure \ref{vmax} and \ref{det_bt}, different definitions of spiral galaxies strongly affect the detection fraction of ALFALFA-SDSS sample. Central spiral galaxies, defined by visual morphology classified by Galaxy Zoo project, achieve a very high \hi\ detection fraction, which is essential for obtaining unbiased intrinsic \hi\ scaling relations. Other definitions result in lower detection fractions. 

As mentioned before, there is a single tight \htwo\ FFR for all galaxies (grey thick line), while both the SFE$_{\rm HI}$-sSFR and $\mu_{\rm HI}$-sSFR relation are systematically dependent on stellar mass for central spiral galaxies, but both relations show a similar slope for different stellar mass bins (i.e., the contour lines in the right two panels of Figure \ref{3d mpa} are nicely parallel to each other). As above, the two relations are closely related. It is important to notice that the \htwo\ FFR has a non-zero slope ($\sim$ 0.5), while the slope of the SFE$_{\rm HI}$-sSFR relation is about unity and that of the $\mu_{\rm HI}$-sSFR relation is about zero for all stellar masses. This indicates that for central spirals, at any stellar mass, when sSFR decreases, both \htwo\ gas mass and SFE$_{\rm H_2}$ decrease. \hi\ gas fraction remains almost constant and SFE$_{\rm HI}$ decreases. The $M_{\rm H_2}/M_{\rm HI}$ ratio also decreases. Previous studies found that there do exist some \hi-rich galaxies with low SFRs \citep[e.g.,][]{2009A&A...498..407G,2012MNRAS.422.1835S,2014ApJ...790...27L,2019ApJ...884L..52Z,2019MNRAS.485.3169P,2020MNRAS.493.1982J,2023MNRAS.526.1573S}. Here we show that for a significant population of central spirals, \hi\ reservoir is ubiquitous, down to $M_* \sim 10^{9}$\Msol.


\section{Summary and Discussion}

Since \hi\ gas usually extends much further than stars, external environmental effects such as interactions and mergers can strongly impact the \hi\ gas. To better understand the processes of in-situ star formation and internal-origin quenching, this study aims to focus exclusively on central galaxies to minimize the strong environmental effects that can significantly influence satellite galaxies. Additionally, a galaxy's visual morphology is more sensitive to external interactions or mergers than structural parameters like the bulge-to-total ratio (B/T), concentration index ($R_{90}$/$R_{50}$), and Sersic index. Therefore, in this work, we focus only on central galaxies those are visually defined as spiral galaxies.

Because both SDSS and ALFALFA are flux-limited surveys, they exhibit strong selection biases in stellar mass and \hi\  gas mass, respectively. Consequently, a joint incompleteness correction is necessary for the SDSS-ALFALFA matched sample. After performing proper incompleteness correction, the average \hi\ detection fraction in the ALFALFA-SDSS matched sample is significantly increased. For the central spiral galaxies (defined by visual morphology), the \hi\ detection fraction is, on average, larger than 90\%, even for many low-mass and low-SFR systems. These are in good consistency with the results from the deeper xGASS survey, hence supporting our incompleteness correction method. It is important to note that if the ``spiral/disk'' are defined by their structural parameters (e.g., bulge-to-total ratio, B/T $<$ 0.3), the average \hi\ detection fraction ($\sim$ 64\%) is significantly lower than that of spiral galaxies defined by visual morphology, especially for galaxies below the SFMS. This is likely because the structure of a galaxy is generally expected to be less sensitive to external environmental effects compared to its visual morphology, which more readily reveals features such as tidal disturbances, mergers, and asymmetries.

The high overall \hi\ detection fraction for the central spiral galaxies (defined by visual morphology) allows for the derivation of the unbiased \hi\ scaling relations for this particular population. We find that the SFE$_{\rm HI}$-sSFR and $\mu_{\rm HI}$-sSFR scaling relations for central spirals show strong systematic dependence on stellar mass. At any given stellar mass, the \hi\ gas mass fraction remains roughly constant with varying sSFR, suggesting that \hi\ reservoir is present in central spirals regardless of their star formation status, down to $M_* \sim 10^{9}$\Msol. 
Along with the tight correlation between the molecular gas mass fraction and sSFR for galaxies across a wide range of different properties, these results indicate that a cessation of \htwo\ supply drives the decline of SFR of central spirals in the local universe, despite the abundance of \hi\ gas. Our findings hence provide key observational evidence of the dramatically different behavior of their multi-phase ISM. 

In regular undisturbed spiral galaxies, the atomic gas usually distributes towards a larger radius than the stellar and molecular gas disk \citep[e.g.,][]{2002ApJ...569..157W,2008AJ....136.2846B,2014ApJ...790...27L,2017A&A...605A..18C}. Gas inflow carrying excess angular momentum can accumulate in a stable outer \hi\ ring, persisting over extended periods in the absence of perturbations \citep{2020MNRAS.491L..51P,2020MNRAS.495L..42R}. This would deplete the material available in the inner disk for new star formation, providing a straightforward explanation for our observations, a phenomenon also supported by simulations \citep[e.g.,][]{2022MNRAS.509.2707L,2022MNRAS.511.6126W}.

Moreover, the surface density of gas at larger radii is often below the critical threshold needed to convert \hi\ into \htwo, necessary for initiating star formation \citep{2008AJ....136.2846B,2008Leroy,2020ARA&A..58..157T,2022ApJ...930...85Y}. Coupled with the tight correlation between molecular gas mass fraction and sSFR across diverse galaxy types, our results indicate that the reduction in SFR in central spirals predominantly results from a cessation in \htwo\ supply and a concomitant decline in \htwo\ star formation efficiency, despite there is plenty of \hi\ gas around. 

Alternatively, the disruption in \htwo\ supply could result from feedback mechanisms. In this scenario, our findings indicate that such feedback should have minimal impact on the \hi\ located in the outer disk regions, thereby imposing significant constraints on the strength and nature of feedback in theoretical models \citep[e.g.,][]{2022ApJ...927..189S,2023ApJ...950L..22W}.

Future high spatial resolution 21 cm observations (e.g., the SKA) for a sample of central spiral galaxies with different star formation levels and feedback status will be necessary to obtain their resolved \hi\ gas properties, including \hi\ surface density, kinematics, and disk instability. Such future 21 cm observations will help determine whether \hi\ in quenched central spirals is confined beyond the optical disk, as proposed in the scenario by \citet{2020MNRAS.491L..51P}, or whether the \hi\ lies within the disk itself but does not convert to \htwo\ for some reason. These observations will be crucial in testing these hypotheses and advancing our knowledge of the complex interplay between gas dynamics, star formation, and feedback processes in galaxies. Dark matter halo may also play an important role in regulating gas cycling and star formation, and will be explored in our following work.  \\


\section*{Acknowledgments}

This work is supported by the National Natural Science Foundation of China (No. 12303010, 12125301, 12192220, 12192222, 12121003, 12192223), and the science research grants from the China Manned Space Project with No. CMS-CSST-2021-A07.

\appendix

\section{Selection Biases in SDSS and ALFALFA samples} \label{appendix a}

\begin{figure*}[htbp]
    \begin{center}
       \includegraphics[width=180mm]{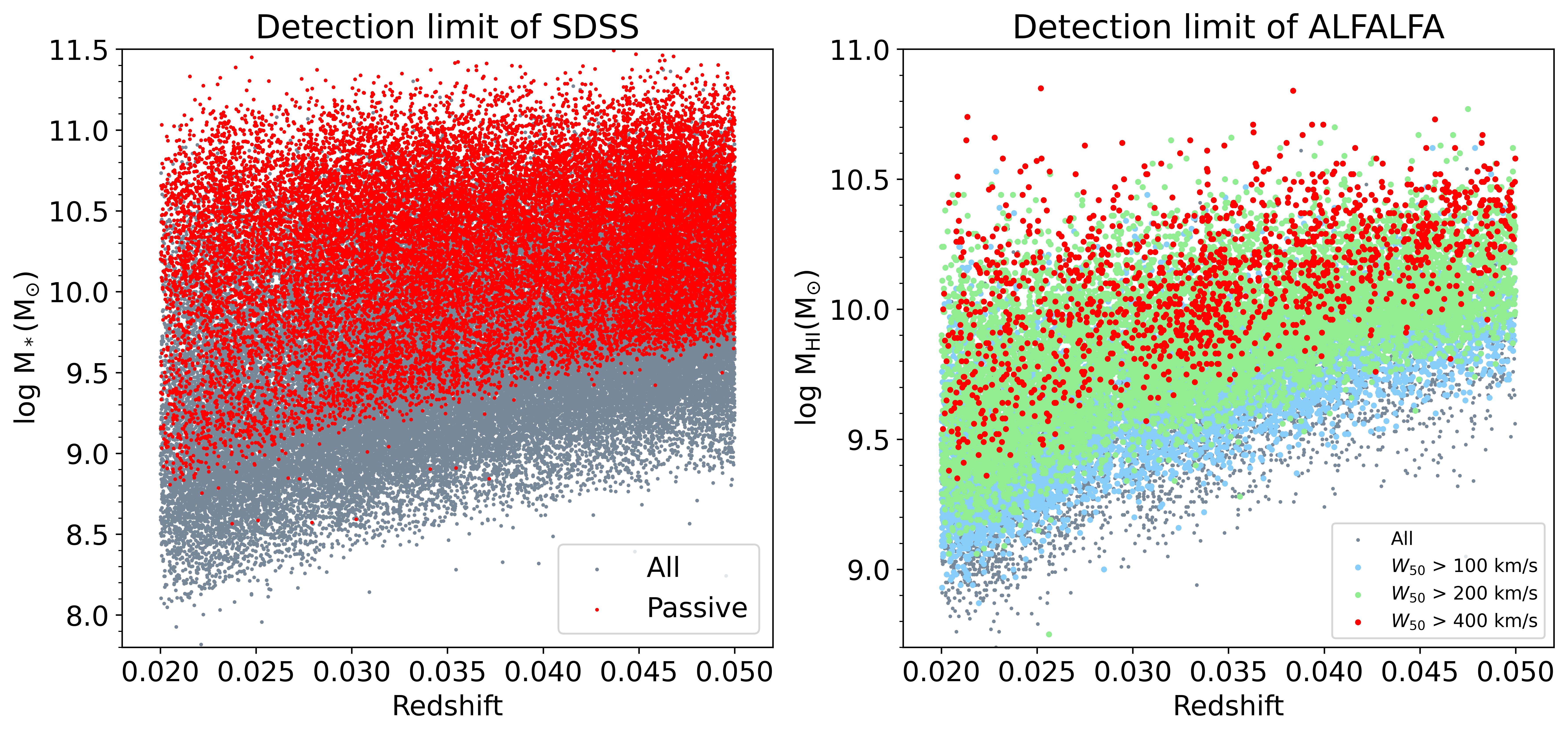}
    \end{center}
\caption{The detection limits of the SDSS spectroscopic and ALFALFA samples. Left: stellar mass ($M_*$) distribution for the SDSS sample in a narrow redshift bin 0.02 $< z <$ 0.05. The galaxies with log sSFR $<$ - 1.8 Gyr$^{-1}$ (the measurements of stellar masses and SFRs are described in section \ref{M and sfr}) are defined to be passive and denoted as red dots. Right: atomic gas mass $M_{\rm HI}$ distribution for the ALFALFA sample in the same redshift bin. The galaxies with $W_{50} >$ 100 km/s, 200 km/s, and 400 km/s are denoted as blue, green and red dots, seperately. Evidently, both SDSS and ALFALFA samples are strongly biased and sample incompleteness corrections are necessary for any statistical analysis. }
 \label{dete_limit}
\end{figure*}

Due to the SDSS spectroscopic selection of $r <$ 17.77, the sample is flux-limited. As shown in the left panel of Figure \ref{dete_limit}, even within the narrow redshift range of 0.02 $<z<$ 0.05, this selection produces a strong bias towards massive and star-forming galaxies, and faint sources are progressively missed at higher redshifts. 

Similar to SDSS, ALFALFA is also a flux-limited sample and the effective integration time is typically 48 seconds for each source, leading to a strong selection in \hi\ gas mass. As shown in \citet{2011AJ....142..170H}, the completeness limits for Code I and Code II sources in ALFALFA are both functions of $W_{50}$, with a break at $W_{50} =$ 300km/s. At the same integrated line flux, ALFALFA sources with narrower \hi\ profiles (i.e., smaller $W_{50}$) host larger S/N, hence are easier to be detected than broader ones. The selection effects of ALFALFA within the narrow redshift range of 0.02 $<z<$ 0.05 are clearly shown in the right panel of Figure \ref{dete_limit}. At a given (low) \hi\ gas mass, the number of \hi\ non-detections increases significantly with both redshift and $W_{50}$ due to sample incompleteness, which must be properly accounted for in any statistical analysis. \\


\section{Comparison between different classifications of spiral galaxies} \label{spiral definitions}

\begin{figure*}[htbp]
    \begin{center}
       \includegraphics[width=180mm]{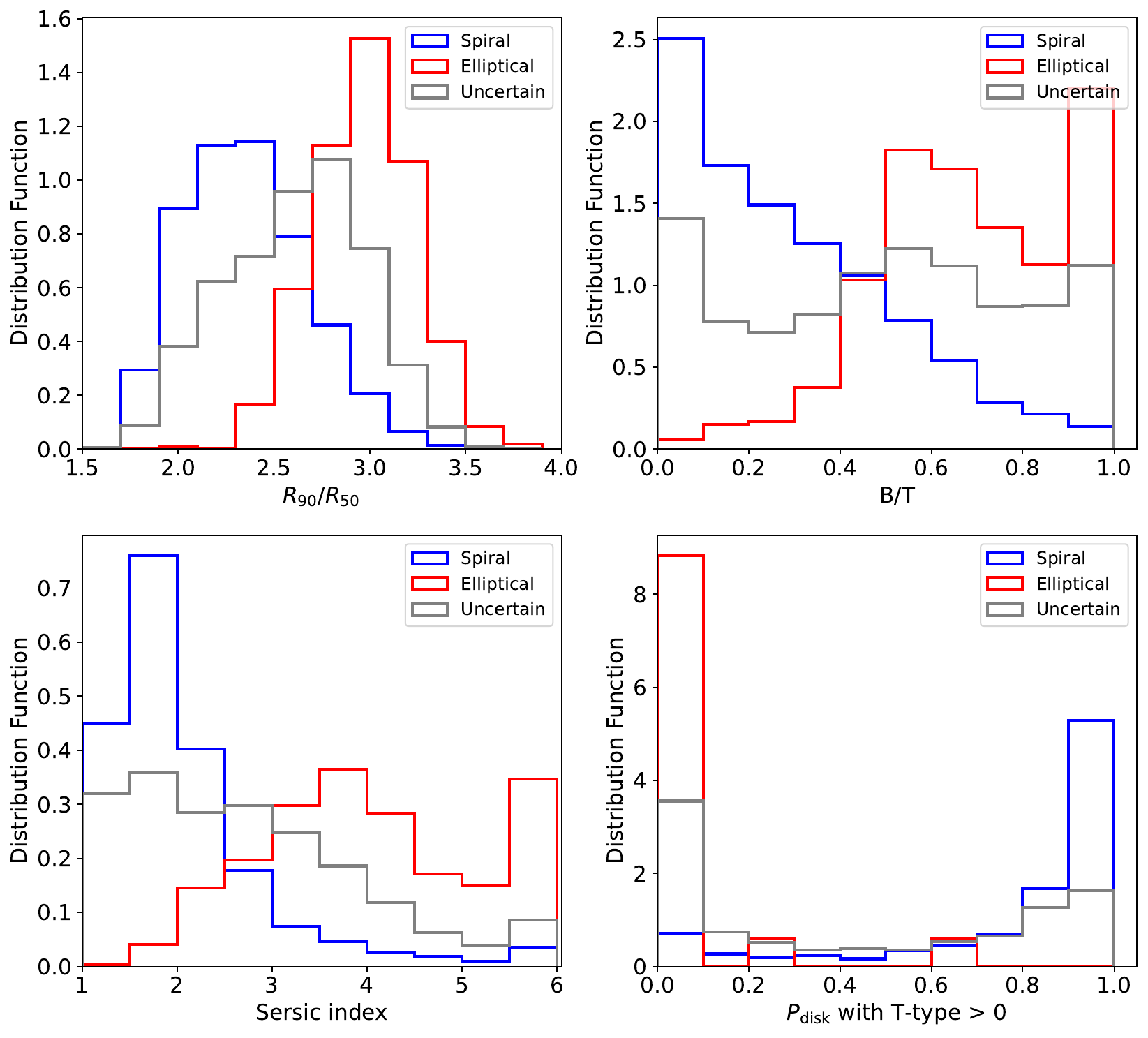}
    \end{center}
\caption{The distribution function of $R_{90}/R_{50}$ in $r$ band (upper left panel), mass-weighted B/T (upper right panel), S\'ersic index in $r$ band (lower left panel) and probability of a galaxy to be late-type disk galaxy with T-type $>$ 0 using the machine learning morphology catalog from \citet{2018MNRAS.476.3661D} (lower right panel) for central galaxies with $10^{10}\Msol < M_* < 10^{10.3}\Msol$ and 0.02 $< z <$ 0.05 in SDSS. In each panel, blue, red, and grey histograms indicate the visual morphological classification of spiral, elliptical, and uncertain galaxies defined from Galaxy Zoo. Clearly, a significant amount of central visually-defined spiral galaxies have large values of $R_{90}/R_{50}$, B/T and S\'ersic index, and will be classified as ``elliptical” according to their structural parameters. On the other hand, many visually classified ``uncertain” galaxies have small values of $R_{90}/R_{50}$, B/T and S\'ersic index, and large value of $P_{\sf disk}$ with T-type $>$ 0, and will be classified as ``spiral” according to their structural parameters or morphology classification from machine learning.}
 \label{structure}
\end{figure*}

Spiral galaxies can be defined by their visual morphology, structure or kinematics. To demonstrate the impact of different methods to define ``spiral" galaxies, in Figure \ref{structure}, we show the distribution function of concentration index $R_{90}/R_{50}$ (upper left panel), bulge-to-total ratios B/T (upper right panel), and S\'ersic index (lower left panel) for central galaxies in a narrow stellar mass range of $10^{10} - 10^{10.3}\Msol$ in the SDSS parent sample within the redshift range 0.02 $< z <$ 0.05. We retrieved the $r$ band S\'ersic index, Petrosian half-light radii $R_{50}$ and $R_{90}$ from the New York University Value-Added Catalog \citep[NYU-VAGC; ][]{2005AJ....129.2562B}. The mass-weighted bulge-to-total ratios (B/T) are taken from \citet{Simard2011} and \citet{2014ApJS..210....3M}, with a fitting model using a pure exponential disk and a de Vaucouleurs bulge. Galaxies are split into ``spirals" (blue histograms, 3166 galaxies), ``ellipticals" (red histograms, 537 galaxies) and ``uncertains" (grey histograms, 2918 galaxies) as classified by Galaxy Zoo (visual classifications). Figure \ref{structure} clearly shows that morphology is not a one-to-one correspondence with structure. Visually-defined spirals/ellipticals/uncertains span a wide range of $R_{90}/R_{50}$, B/T, and S\'ersic index. The results for other stellar mass bins also show similar results that at a given structural parameter, there are significant overlaps between the three populations. Hence, different definitions (i.e., defined by visual morphology or structural parameters) can produce significantly different galaxy samples.

It should be noted that the machine learning (ML) method trained on visually classified morphology (e.g., Galaxy Zoo) has also been used widely to classify galaxies. For instance, using an ML-based morphology catalog from \citet{2018MNRAS.476.3661D}, disk galaxies can be selected with T-type $>$ 0 and $P_{\sf disk} >$ 0.5, where T-type describes the galaxy morphology type and galaxies with T-type $>$ 0 correspond to late-type morphologies in the Hubble sequence, and $P_{\sf disk}$ is the probability for a galaxy to be a disk. The lower right panel of Figure \ref{structure}  clearly shows that for the centrals with T-type $>$ 0 and $P_{\sf disk} >$ 0.5, there are significant numbers of ``uncertains” as classified in Galaxy Zoo. Therefore, although the ML-based morphology classification works very well in general for all galaxy populations in SDSS, at least for this particular case, it badly failed to recover the human eyeball classified results. As discussed before, many disk-like ``uncertains" share similar structure parameters (e.g., B/T, sersic index) as ``spirals", but their visual morphology features look different from ``spirals". These differences maybe related to external environmental effects such as perturbation and interaction, which are difficult to be catched by some of the ML method. These external environmental effects tend to produce \hi-poor systems. \\

\section{Bias introduced by 1-dimensional statistical analysis}  \label{bias}

\begin{figure*}[htbp]
    \begin{center}
       \includegraphics[width=180mm]{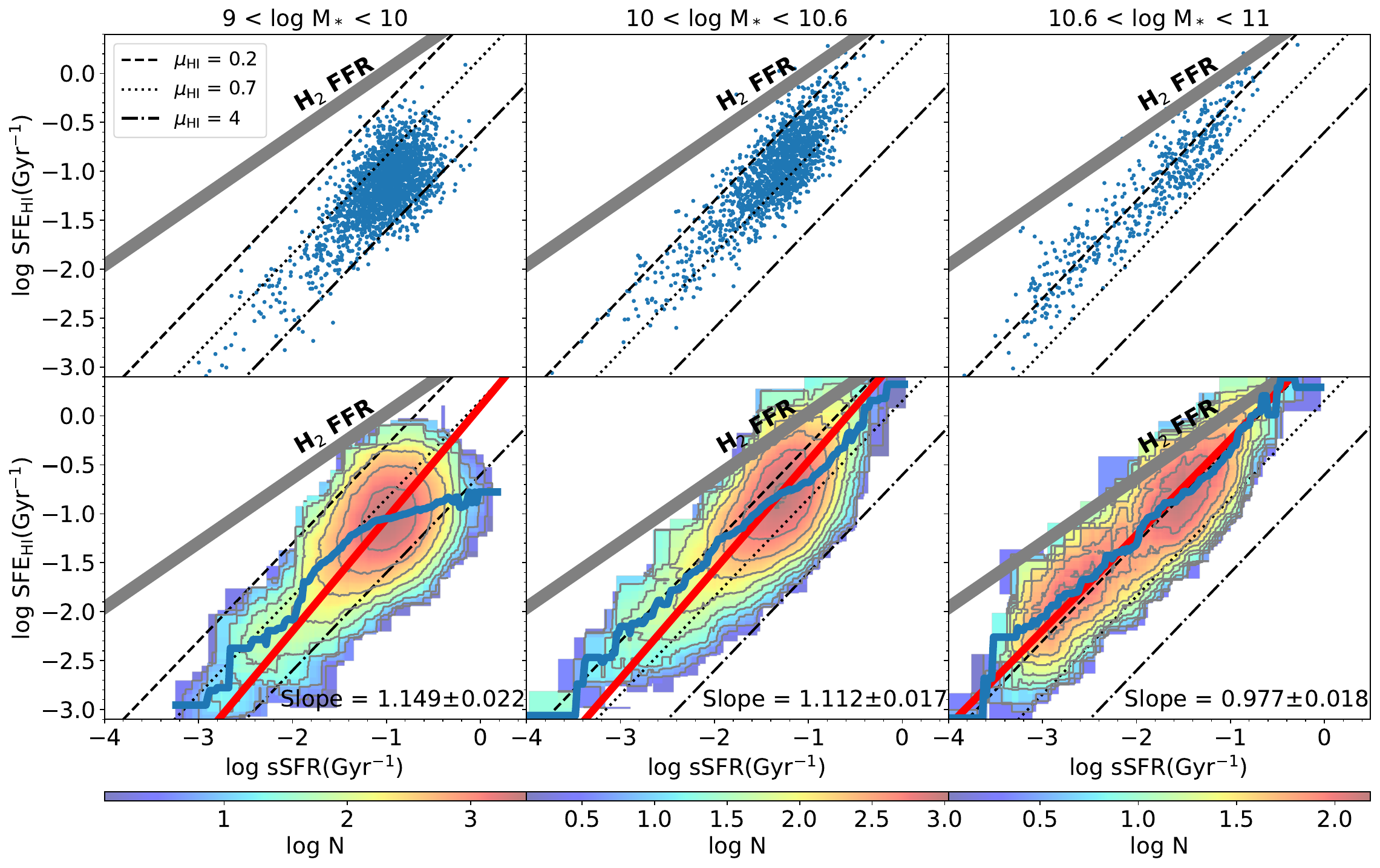}
    \end{center}
\caption{The distribution of central spiral galaxies in the ALFALFA-SDSS sample on the SFE$_{\rm HI}$-sSFR plane, splitted into three different stellar mass bins. In each panel, the black diagonal lines indicate three constant $\mu_{\rm HI}$. The grey thick lines indicate the molecular gas FFR for references. Blue dots in the upper three panels are individual galaxies. Level contours for the number density of galaxies are shown in the lower three panels, with colors ranging from blue to red as a function of logarithm of the number density, applied with sample incompleteness correction. The blue lines show the average SFE$_{\rm HI}$ (at a given sSFR) as a function of sSFR, calculated with a moving average of 0.5 dex in sSFR. The red lines show the best fits to the data using the ODR fitting method, with the slope given in the legend, which is around unity (i.e., galaxies distribute following the diagonal constant \hi\ gas-to-mass ratio line) for all three mass bins. While the blue line has, on average, a shallower slope than unity, indicating that when sSFR decreases, the \hi\ gas-to-mass ratio also decreases, i.e., galaxies become more \hi-poor. Evidently, the red line better traces the ridge of the contour lines than the blue line, and hence the ODR fitting that considers the variations in both the x- and y-axes represents a more accurate fit to the data, in particular for the case of low mass bin where the scatter is larger.  }
 \label{map test}
\end{figure*}

Figure \ref{map test} shows the SFE$_{\rm HI}$-sSFR relation of central spirals in the ALFALFA-SDSS matched sample in three stellar mass bins. Blue dots in the upper panels are individual galaxies. Level contours for the number density of the galaxies at given stellar mass bins are shown in the lower three panels, with colors ranging from blue to red as a function of number density, applied with sample incompleteness corrections. The red lines are the best fits to the data using the orthogonal distance regression (ODR) fitting method, which calculates the sum of the orthogonal distances from the data points to the fitted line. The blue lines show the average SFE$_{\rm HI}$ (at a given sSFR) as a function of sSFR, calculated with a moving average of 0.5 dex in sSFR.

In the lower panels, the contour lines of the 2-dimensional density distributions clearly indicate the underlying trends. Evidently, the ODR fitting (red line) better traces the ridge of the contour lines than the blue line, and hence represents a more accurate fit to the data, in particular for the case of low mass bin where the scatter is larger. This is expected since ODR fitting takes the variation in both x- and y-axes into account. The slope of the best ODR fits is around unity in all three stellar mass bins, indicating that galaxies distribute along with the constant gas-to-mass ratio, i.e., when sSFR decreases, the \hi\ gas-to-mass ratio remains constant. 

While the commonly used 1-dimensional scaling relation (e.g., the blue lines), which considers only the variation in the y-axis at a given x, agrees with the ridge line and ODR fitting for the massive galaxies (lower right panel), due to the relatively narrow data distribution, i.e., the variation is small in both x- and y-axes. However, in the low-mass bin (lower left panel), the blue line significantly deviates from the ridge line and the red line, due to the larger scatter and the shape of the data distribution. Also, the blue line has, on average, a shallower slope than unity, indicating that when sSFR decreases, the \hi\ gas-to-mass ratio also decreases, i.e., galaxies become more \hi-poor. 

The results shown in Figure \ref{map test} demonstrate that using 1-dimensional scaling relations (e.g., the blue lines in the lower panels in Figure \ref{map test} and the lines in the middle panels in Figure \ref{3d mpa} to describe the trend can be biased, due to the shape of the data distribution. For instance, in the upper middle panel in Figure \ref{3d mpa}, the fact that the blue line (lowest stellar mass bin) is flattened out at high sSFR is due to the effect described in Figure \ref{map test}. Clearly, the corresponding contour lines in the upper left panel are all well parallel to the diagonal constant \hi\ gas-to-mass ratio line. \\

\section{Alternative SFR indicators}  \label{sfr indicators}

\begin{figure*}[htbp]
    \begin{center}
       \includegraphics[width=180mm]{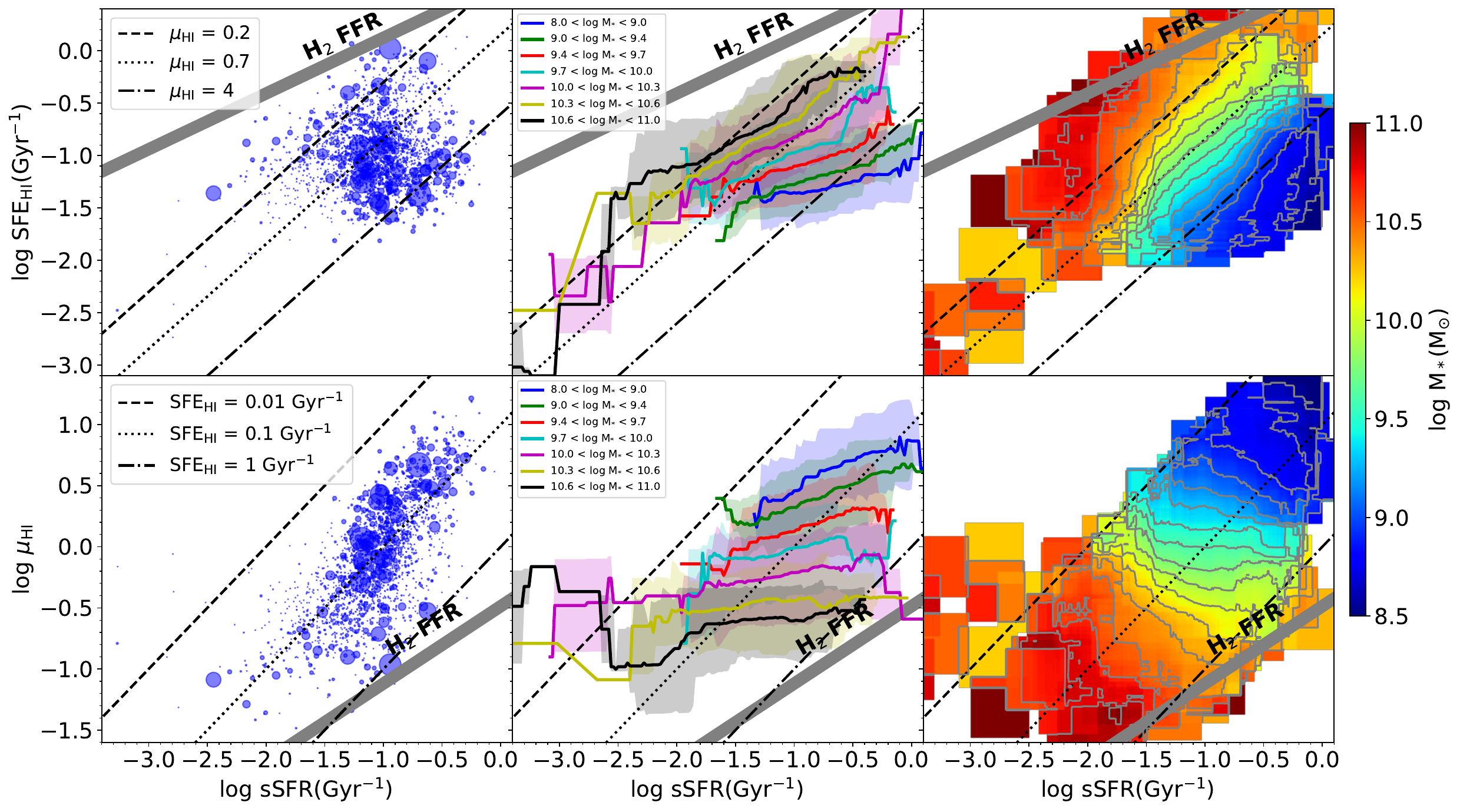}
    \end{center}
\caption{As for Figure \ref{3d mpa}, but using SFRs derived from the SED fitting of UV, optical and mid-IR bands \citep{2018ApJ...859...11S}. The \htwo\ FFR is also derived from the SED SFRs, correspondingly. The range of the galaxy distribution in sSFR is narrower compared to Figure \ref{3d mpa}, but the general trends shown in the right two panels (which are more accurate representations of the true underlying trends compared to the 1-dimensional analysis shown in the middle two panels, as illustrated in Figure \ref{map test}) are similar to those in Figure \ref{3d mpa}.}
 \label{salim}
\end{figure*}

As discussed in the main text, with proper sample incompleteness corrections, for central spirals, the overall \hi\ detection DF reaches $\sim$ 95\% for the ALFALFA-SDSS matched sample, and $\sim$ 98\% for the xGASS sample. Hence, our main conclusion that there is a ubiquitous \hi\ reservoir in the vast majority of central spirals (defined by visual morphology) is largely independent of any particular SFR indicator. Here we further test the detailed \hi\ scaling relations shown in Figure \ref{3d mpa} with alternative SFR indicators. 

We repeat the analyses using SED SFRs derived from UV, optical and mid-IR fluxes \citep[GSWLC-M2;][]{2016ApJS..227....2S,2018ApJ...859...11S}. The results are shown in Figure \ref{salim}. As discussed in the previous Section and Figure \ref{map test}, the trends derived from the 1-dimensional method (the middle two panels of Figure \ref{salim}) can be significantly biased. Using 2-dimensional analysis and ODR fitting produces more accurate representations of the true underlying trends. Therefore, comparing the right two panels of Figure \ref{salim} with those in Figure \ref{3d mpa}, although the range of the galaxy distribution in sSFR is narrower (due to the use of different SFR indicators), the general trends are similar. Both SFR indicators reveal that the SFE$_{\rm HI}$-sSFR relation and $\mu_{\rm HI}$-sSFR relation strongly correlate with stellar mass, and that at a given stellar mass, $\mu_{\rm HI}$ is largely constant across sSFR. 

It should be noted that although there are few fully quenched central spirals with log sSFR $<$ -2 Gyr$^{-1}$ based on the SED-based SFRs, our main conclusions that there is a ubiquitous \hi\ reservoir in the vast majority of central spirals remain robust, and that at a given stellar mass, $\mu_{\rm HI}$ is largely constant across sSFR. The exact name of the galaxies with the lowest sSFR, i.e., fully quenched or green valley, is therefore less critical. 

In addition, in the right two panels in Figure \ref{salim}, we also show the \htwo\ FFR derived from the SED-based SFRs, which has a steeper slope than the one in Figure \ref{3d mpa}. It is important to note that though the amount of the decrease in sSFR is smaller for the SED-based SFRs (due to the narrower range of sSFR), the corresponded amount of decrease in the \htwo\ gas mass is similar to those shown in Figure \ref{3d mpa}. The $\sim$10 times less \htwo\ gas accompanied by the suppressed star formation efficiency in the central spirals with the lowest sSFRs indicates that they are indeed in the process of being quenched, while their \hi\ gas mass fraction remains about constant.\\


\end{document}